\documentclass[11pt,authoryear,preprint]{elsarticle}

\usepackage{color}

\begin{document}

\title{Integration of the Euler-Poinsot Problem in New Variables}

\author[roa]{Martin Lara\corref{cor1}}
\ead{mlara@roa.es}
\address[roa]{Real Observatorio de la Armada, 11110 San Fernando, Spain}

\author[um]{Sebasti\'an Ferrer}
\ead{sferrer@um.es}
\address[um]{Departamento de Matem\'atica Aplicada, Universidad de Murcia, 30100 Murcia, Spain}

\cortext[cor1]{Corresponding author}

\journal{Mechanics Research Communications}

\begin{abstract}
The essentially unique reduction of the Euler-Poinsot problem may be performed in different sets of variables. Action-angle variables are usually preferred because of their suitability for approaching perturbed rigid-body motion. But they are just one among the variety of sets of canonical coordinates that integrate the problem. We present an alternate set of variables that, while allowing for similar performances than action-angles in the study of perturbed problems, show an important advantage over them: Their transformation from and to Andoyer variables is given in explicit form.
\end{abstract}

\maketitle

\section{Introduction}

The Euler-Poinsot problem is a three degrees-of-freedom (DOF) problem whose super-integrable character limits the solutions to quasi-periodic orbits on two-torus \citep{Fasso2005}. Because of the symmetry with respect to rotations about the angular momentum vector, the problem is formulated as a 1-DOF Hamiltonian when using \citet{Andoyer1923} variables. Then, the essentially unique complete reduction that provides the integration of the problem can be performed in different variables, and it is usually done by solving the Hamilton-Jacobi equation.
\par

In the study of the rigid body rotation under external torques the use of suitable variables reveals crucial to the solution by perturbation methods. A common trend is to use action-angle variables \citep{Sadov1970, SadovRuso1970,Kinoshita1972}, but other variables can be used instead \citep{HitzlBreakwell1971}.
\par

The Hamilton-Jacobi equation of the Euler-Poinsot problem in Andoyer variables can be solved formally, without need of specifying the new Hamiltonian \citep{FerrerLara2010}. We show how this new, unspecified Hamiltonian can be cast into a standard form in which the modulus of the elliptic integrals that appear in the solution of the transformation, remains as an undetermined state function of the new momenta. Under certain conditions imposed to the formal transformation, the modulus is determined by solving a system of partial differential equations. The condition we require is ``simplification'' and find a new set of variables that while showing similar performances than action-angles \citep{Sadov1970,SadovRuso1970}, has the benefit of not requiring the computation of implicit functions. In addition, we demonstrate that Sadov's transformation is also a member of the general family of Euler-Poinsot transformations to Andoyer variables.

\section{Complete Reduction of the Euler-Poinsot Problem}

The Hamiltonian of the torque-free rotation is \citep{DepritAJP1967}
\begin{equation}\label{Andoyer}
\mathcal{H}=\left(\sin^2\nu/A+\cos^2\nu/B\right)(M^2-N^2)/2+N^2/2C,
\end{equation}
where $A$, $B$, and $C$ are the principal moments of inertia of the body, and the Andoyer variables are defined by three pairs of conjugate variables: the rotation angle on the equatorial plane of the body $\nu$ and the projection of the angular momentum vector on the body axis of maxima inertia $N$, the precession angle on the invariant plane $\mu$ and the modulus of the angular momentum vector $M$, and the node angle on the inertial plane $\lambda$ and the projection of the angular momentum vector on the axis perpendicular to the inertial plane $\Lambda$. Because $\lambda$, $\Lambda$ and $\mu$ are cyclic $\lambda=\lambda_0$, $\Lambda=\Lambda_0$, and $M=M_0$ are constant, and Eq. (\ref{Andoyer}) is a Hamiltonian of 1-DOF.
\par

The integration may be done by complete reduction. To this goal, we look for canonical transformations
$\mathcal{T}_{\mathcal{K}}:(\lambda,\mu,\nu,\Lambda,M,N)\rightarrow(\ell,{g},{h},L,{G},{H})$
that convert Eq. (\ref{Andoyer}) in a new Hamiltonian $\mathcal{K}$ that depends only on momenta. Because of the two-torus topology of the Euler-Poinsot problem only two momenta are required in $\mathcal{K}$, and in view of neither $\lambda$ nor $\Lambda$ appear in Eq. (\ref{Andoyer}), we choose $h=\lambda$, $H=\Lambda$ and $\mathcal{K}\equiv\mathcal{K}(L,G)$.
\par

\subsection{Formal Solution of the Hamilton-Jacobi Equation}

In the Hamilton-Jacobi approach, the transformation $\mathcal{T}_{\mathcal{K}}$ is derived from a generating function in mixed variables $\mathcal{S}=\mathcal{S}(\mu,\nu,L,G)$ such that
\begin{equation} \label{transformation0}
(\ell,g,M,N)=\frac{\partial{S}}{\partial(L,G,\mu,\nu)}
\end{equation}
Because $\mu$ is cyclic in Eq. (\ref{Andoyer}), $\mathcal{S}$ is chosen in separate variables
$\mathcal{S}=G\,\mu+W(\nu,L,G)$. Then, from Eq. (\ref{Andoyer}) we form the Hamilton-Jacobi equation
\begin{equation}\label{HJsolido}
\left(\frac{\sin^2\nu}{2A}+\frac{\cos^2\nu}{2B}\right)\left[G^2-\left(\frac{\partial{W}}{\partial\nu}\right)^{\!\!2}\right]+\frac{1}{2C}\left(\frac{\partial{W}}{\partial\nu}\right)^{\!\!2}=\mathcal{K}
\end{equation}
where $W$ may be solved by quadrature. Then, calling $\beta=L/G$, the transformation Eqs. (\ref{transformation0}) are
\begin{eqnarray} 
\label{delta}
\ell &=& \frac{\mathcal{I}_2}{G^2}\,\frac{\partial\mathcal{K}}{\partial\beta},
\\ \label{gamma}
g &=& \mu+\mathcal{I}_1-\frac{\mathcal{I}_2}{G^2}\left(2\mathcal{K}
+ \beta\,\frac{\partial\mathcal{K}}{\partial\beta}\right),
\\ \label{N}
N &=& G\,\sqrt{Q}, 
\\ \label{M}
M &=& G,
\end{eqnarray}
where
\begin{equation}\label{II12}
\mathcal{I}_1=\int_{\nu_0}^\nu\sqrt{Q}\,\mathrm{d}\nu, \qquad
\mathcal{I}_2=\int_{\nu_0}^\nu\frac{1}{\sqrt{Q}}\,\frac{\partial{Q}}{\partial(1/\Delta)}\,\mathrm{d}\nu,
\end{equation}
\begin{equation}
Q=\frac{\sin^2\nu/A+\cos^2\nu/B-1/\Delta}{\sin^2\nu/A+\cos^2\nu/B-1/C},
\end{equation}
and
\begin{equation}\label{K}
1/\Delta=2\mathcal{K}/G^2.
\end{equation}
\par

We only discuss the general case $A<B<C$. As $\sqrt{Q}$ must be real for all $\nu$, we get $B\le\Delta\le{C}$ and $\frac{1}{2}{G}^2/C\le\mathcal{K}\le\frac{1}{2}{G}^2/B$, thus constraining the motion to rotations about the axis of maxima inertia. The discussion of other cases is left to the reader.
\par

The transformation equations for $\ell$ and $g$, Eqs. (\ref{delta})--(\ref{gamma}), depend on the integration of the two quadratures in Eq. (\ref{II12}). However, as far as $\mathcal{K}$ depends only on the momenta $G$ and $L$, the quadratures in Eq. (\ref{II12}) can be solved without need of specifying the formal dependence of $\mathcal{K}$ on the new momenta, thereby giving rise to a whole family of canonical transformations \citep{FerrerLara2010}.
\par

The closed form solution of Eq. (\ref{II12}) relies on well known changes of var\-i\-a\-bles. Thus, introducing the parameter $f>0$ and the function $0\le{m}\le1$
\begin{eqnarray}
\label{efe}
f=\frac{C\,(B-A)}{(C-B)\,A},\qquad
m=\frac{(C-\Delta)\,(B-A)}{(C-B)\,(\Delta-A)},
\end{eqnarray}
and the auxiliary variable $\psi$ defined as
\begin{equation}\label{psi2nu}
\cos\psi=\frac{\sqrt{1+f}\sin\nu}{\sqrt{1+f\sin^2\nu}},\qquad
\sin\psi=\frac{\cos\nu}{\sqrt{1+f\sin^2\nu}},
\end{equation}
then, the quadratures in Eq. (\ref{II12}) are solved to give
\begin{eqnarray}
\mathcal{I}_1 &=& \gamma\left[\frac{m}{f+m}\,F(\psi|m)-\Pi(-f,\psi\,|\,m)\right],
\\[1ex]
\mathcal{I}_2 &=& \gamma\,\frac{A\,C}{C-A}\,F(\psi|m),
\end{eqnarray}
where
\begin{equation}\label{chi}
\gamma=\sqrt{(1+f)\,(f+m)/f}=\sqrt{\frac{B\,\Delta\,(C-A)\,(C-A)}{A\,C\,(C-B)\,(\Delta-A)}},
\end{equation}
$F(\psi|m)$ is the elliptic integral of the first kind of modulus $m$ and amplitude $\psi$, and $\Pi(-f,\psi\,|\,m)$ is the elliptic integral of the third kind of modulus $m$, amplitude $\psi$, and characteristic $-f$. It is worth mentioning that for the definition of the elliptic integral of third kind we adhere to the convention in \citep{ByrdFriedman1971}.
\par

\subsection{The Standard Hamiltonian}
From Eqs. (\ref{K}) and (\ref{efe}) we note that $\mathcal{K}$ is characterized by the identity
\begin{equation}\label{SadovHam}
\mathcal{K}=\frac{G^2}{2A}\left(1-\frac{C-A}{C}\frac{f}{f+m}\right),
\end{equation}
which can be taken as a definition by assuming that $m=m(L,G)$ in Eq. (\ref{SadovHam}). Then, Eqs. (\ref{delta})--(\ref{M}) are rewritten
\begin{eqnarray}\label{lg}
\ell &=& 
\frac{1}{2\gamma}\,\frac{1+f}{f+m}\,\frac{\partial{m}}{\partial\beta}\,F(\psi|m),\\ \label{gg}
g &=& 
\mu+\gamma\left[\frac{1}{f+m}\left(m-\frac{f}{f+m}\,\frac{\beta}{2}\,\frac{\partial{m}}{\partial\beta}\right)F(\psi|m)-\Pi(-f,\psi|m)\right]\!,\quad
\\ 
N &=& G\,\sqrt{\frac{f}{f+m}}\,\sqrt{1-m\sin^2\psi}, \\ \label{Mg}
M &=& G.
\end{eqnarray}
\par

Transformations in the literature can be obtained as particular cases of the general form Eqs. (\ref{SadovHam})--(\ref{Mg}). Thus, the new Hamiltonian selected by \citet{HitzlBreakwell1971} is the average of the Andoyer Hamiltonian Eq. (\ref{Andoyer}), which is also the intermediate Hamiltonian of \citet{Kinoshita1972}, while a previous proposal of ours \citep{FerrerLara2010} transforms the Andoyer Hamiltonian to the axisymmetric case.
\par

\section{New variables}

Searching for simplification in Eqs. (\ref{lg}) and (\ref{gg}) we propose to choose
\begin{equation}\label{simplify}
\frac{1}{2\gamma}\,\frac{1+f}{f+m}\,\frac{\partial{m}}{\partial\beta}=-1,\qquad
\frac{1}{f+m}\left(m-\frac{f}{f+m}\,\frac{\beta}{2}\,\frac{\partial{m}}{\partial\beta}\right)=1.
\end{equation}
Equations (\ref{simplify}) can be solved for $\beta=\beta(m)$ without need of solving any partial differential equation. Furthermore, by squaring $\beta$ we can express $m$ as a function of $L/G$
\begin{equation}\label{mmia}
m=f\left[(1+f)\,G^2/L^2-1\right]\!.
\end{equation}
Correspondingly, the new Hamiltonian in new variables is
\begin{equation}\label{newH}
\mathcal{K}=\frac{G^2}{2A}-\left(\frac{1}{B}-\frac{1}{C}\right)\frac{L^2}{2},
\end{equation}
whose Hessian never vanishes, and which is formally equal to the uniaxial case for a new maximum momentum of inertia, say $P$, such that $1/P=1/A+1/C-1/B$.
\par

Then, the direct transformation is
\begin{eqnarray}\label{ell}
\ell &=& -F(\psi\,|\,m) \\
g &=& \mu+\sqrt{(1+f)\,(f+m)/f}\,\left[F(\psi\,|\,m)-\Pi(-f,\psi\,|\,m)\right] \\
L &=& N\,\sqrt{(1+f)/(1-m\sin^2\psi)} \\ \label{H}
G &=& M
\end{eqnarray}
where $\psi$ is defined in Eq. (\ref{psi2nu}) and $m$ is easily computed in Andoyer variables from its definition in Eq. (\ref{efe}) by noting that $\Delta=G^2/(2\mathcal{K})=M^2/(2\mathcal{H})$, where $\mathcal{H}$ is given in Eq. (\ref{Andoyer}).
\par

The inverse transformation requires using the Jacobi amplitude $\mathrm{am}$ to invert Eq. (\ref{ell})
\begin{equation}\label{psi}
\psi=-\mathrm{am}\left(\ell\,|\,m\right),
\end{equation}
where $m$ is computed from Eq. (\ref{mmia}).
Then, from Eq. (\ref{psi2nu}) we get
\begin{equation}\label{Deprit2Serretnu}
\cos\nu=-\frac{\sqrt{1+f}\,\mathrm{sn}(\ell,m)}{\sqrt{1+f\,\mathrm{sn}^2(\ell,m)}},\qquad
\sin\nu=\frac{\mathrm{cn}(\ell,m)}{\sqrt{1+f\,\mathrm{sn}^2(\ell,m)}}.
\end{equation}
where $\mathrm{sn}$, $\mathrm{cn}$, $\mathrm{dn}$, stand for the usual Jacobi elliptic functions.
Finally, the inverse transformation of Eqs. (\ref{ell})--(\ref{H}) is completed with 
\begin{eqnarray} \label{mu}
\mu &=& g+(1+f)\,(G/L)\left[\ell+\Pi(-f,\mathrm{am}\left(\ell\,|\,m\right)\,|\,m)\right], \\ \label{ene}
N &=& L\,\mathrm{dn}(\ell,m)/\sqrt{1+f}, \\ \label{EME}
M &=& G.
\end{eqnarray}

\section{Transformation to action-angle variables}

Note in Eq. (\ref{psi}) that the variable $\ell$ is $4\,K(m)$-periodic, with $K(m)$ the complete elliptic integral of the first kind. With a view on perturbations of the Euler-Poinsot problem, where elliptic functions would be expanded in Fourier series, it could be desired that $\ell$ be $2\pi$-periodic (an angle).
\par

The new variable
\begin{equation}\label{ell2}
\ell'=-\frac{\pi}{2K(m)}\,F(\psi|m),
\end{equation}
will be obtained by requiring to Eq. (\ref{lg}) that
\begin{equation}\label{aacondition}
\frac{1}{2\gamma}\,\frac{1+f}{f+m}\,\frac{\partial{m}}{\partial\beta'}=-\frac{\pi}{2K(m)},
\end{equation}
where $\beta'=L'/G'$. Equation (\ref{aacondition}) is in separate variables and is integrated by quadrature to give, up to an integration constant,
\begin{equation}\label{beta3}
\beta'=\frac{2}{\pi}\,\sqrt{(1+f)\,(f+m)/f}\left[\Pi(-f,m)-\frac{m}{f+m}\,K(m)\right],
\end{equation}
where $\Pi(-f,m)$ is the complete elliptic integral of the third kind.
Equation (\ref{beta3}) defines $m$ as implicit function of $L'/G'$.
\par

Now, we replace $\beta'$ in Eq. (\ref{gg}) by its value from Eq. (\ref{beta3}) to get
\begin{equation}\label{ge2}
g'=\mu+\sqrt{(1+f)\,(f+m)/f}\left[\frac{\Pi(-f,m)}{K(m)}\,F(\psi|m)-\Pi(-f,\psi|m)\right]\!.
\end{equation}

Remarkably, Eqs. (\ref{Mg}), (\ref{ell2}), (\ref{beta3}) and (\ref{ge2}) recover the original transformation to action-angle variables \citep{Sadov1970,SadovRuso1970} without need of relying on their classical definition.\footnote{%
In \citep{Sadov1970,SadovRuso1970}, $f\equiv\kappa^2$ and $m\equiv\lambda^2$, Andoyer variables are $(h,\psi,\phi,L,G,G_\zeta)\equiv(\lambda,\mu,\nu,\Lambda,M,N) $, and the action-angles are $(f,\nu,h,I,G,L)\equiv(\ell',g',h',L',G',H')$. Besides, Sadov's auxiliary angle is $\xi=-\psi$. Note that there is a typo in the definition of $\lambda^2$ in Eq. (4) of \citep{Sadov1970}, which should be multiplied by  $A/C$. This typo is easily traced in Eq. (2.20) of \citep{SadovRuso1970}, but it still remains in \citep{SadovRuso1984,Kozlov2000}.
}

Finally, we note that the inverse transformation from action-angles to Andoyer variables requires the computation of $m$ from the implicit function Eq. (\ref{beta3}).

\section{Conclusions}

Action-angle variables are not necessarily the better option for dealing with perturbed motion. This fact is very well known for perturbed Kep\-ler\-i\-an motion where Delaunay variables are used customarily. The same happens to rotational motion where action-angle variables have the inconvenience of being related to Andoyer variables through implicit relations. But the complete reduction of the Euler-Poinsot problem may be achieved in a variety of canonical variables. Indeed, we demonstrate that when solving the Hamilton-Jacobi equation of the Euler-Poinsot problem in Andoyer variables, the new Hamiltonian can be cast into a standard form as a function of the modulus of the elliptic integrals required in the solution, a quantity that is consubstantial to the problem. Then, the solution of the Hamilton-Hacobi equation can be written formally as a function of the modulus and its partial derivatives with respect to the new momenta. Solving these partial derivatives according to certain criteria provides the desired transformation. In our case, we require ``simplification'' and find a new transformation of variables that, while having similar characteristics than action-angles variables, does not rely on implicit functions. Besides, we show that the transformation to action-angle variables pertains also to this general family.

\section{Acknowledgemnts}

We thank support from the Spanish Ministry of Science and Innovation, pro\-jects
AYA 2009-11896 (M.L.) and MTM 2009-10767 (S.F.),  and from Fundaci\'on S\'eneca of the autonomous region of Murcia (grant 12006/PI/09). We are indebt with Prof.~Sadov, Russian Academy of Sciences, for sending us copies of his preprints in Russian.

\end{document}